\documentclass[intlimits,twoside,a4paper]{article}

\usepackage{amsmath,amssymb}
\usepackage{graphicx}
\usepackage{wrapfig}

\usepackage[T2A]{fontenc}
\usepackage[cp1251]{inputenc}

\usepackage[eqsecnum]{cmpj2}

\issue{2014}{17}{2}{25001}
\doinumber{10.5488/CMP.17.25001}

\articletype{Comments and Replies}

\title[Discussion on Condens. Matter Phys., 2010, Vol.~13, 13703]%
{Discussion on ``Characterization of 1--3 piezoelectric polymer composites~--- a
numerical and analytical evaluation procedure for thickness mode vibrations''
by C.V. Madhusudhana Rao, G. Prasad, Condensed Matter Physics, 2010, Vol.~13, No.~1, 13703%
}
\author[Y. Sun, Z. Li]{Y. Sun\thanks{E-mail: sunyang@pku.edu.cn}\, , \ Z. Li}
\address{School of Electronics Engineering and Computer Science, Peking University, \\
No.5 Yiheyuan Road Haidian District, Beijing, 100871, China}

\date{Received April 4, 2014, in final form May 5, 2014}
\authorcopyright{Y. Sun, Z. Li, 2014}

\begin{document}

\maketitle

\begin{abstract}
In the paper entitled ``Characterization of 1--3 piezoelectric polymer composites~--- a numerical and analytical
evaluation procedure for thickness mode vibrations'', the dependence of the thickness electromechanical coupling
coefficient on the aspect ratio of piezoceramic fibers is studied by finite element simulation for various
volume fractions of piezoceramic fibers in a 1--3 composite. The accuracy of the results is questionable because
the boundary condition claiming that `predefined displacements are applied perpendicularly on $C^+$ plane on all nodes' is not
suitable for the analysis of 1--3 composite with comparatively large aspect ratio from 0.2 to 1. A discussion regarding
this problem and the suggested corrections are presented in this paper.
\pacs 77.65.-j, 77.84.-s, 77.65.-j, 81.05.Qk, 02.70.Dh
\end{abstract}

\section{Problem in the original paper}

\begin{figure}[!t]
\centerline{\includegraphics[width=0.65\textwidth]{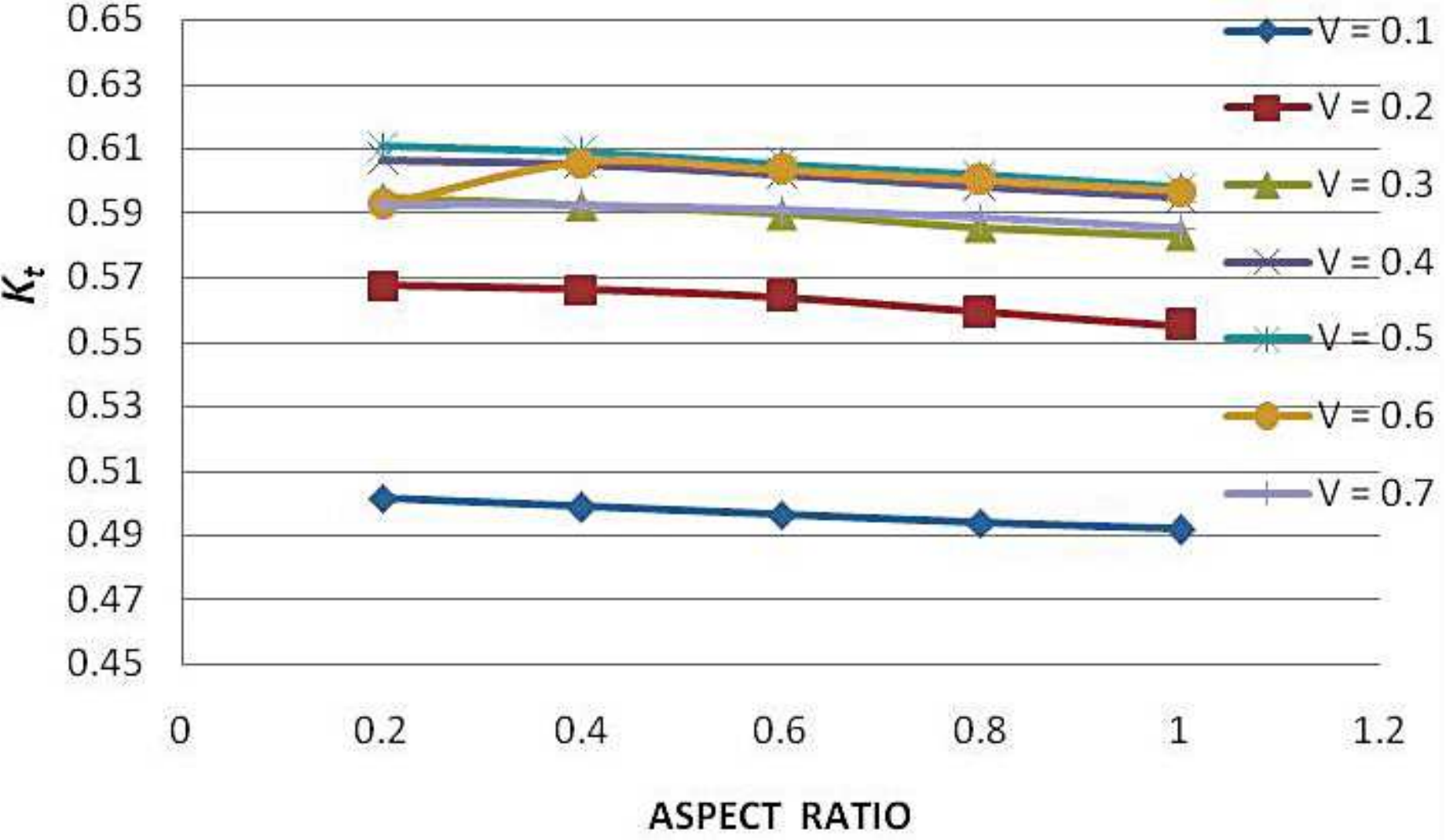}}
\caption{(Color online) Variation of $k_\textrm{t}$ with aspect ratio (from reference \cite{Mad10}).} \label{figure-1}
\end{figure}

In C.V. Madhusudhana Rao's and G. Prasad's paper entitled ``Characterization of 1--3 piezoelectric polymer
composites~--- a numerical and analytical evaluation procedure for thickness mode vibrations'', the dependence
of the thickness electromechanical coupling coefficient $k_\textrm{t}$ on the aspect ratio of piezoceramic
fibers is studied by finite element simulation for various volume fractions of piezoceramic fibers in
the 1--3 composite, and the results are shown in figure~\ref{figure-1} (figure~9 in reference~\cite{Mad10}).
It can be seen that for every fixed volume fraction from 10\% to 70\%, the electromechanical coupling
coefficient $k_\textrm{t}$ decreases with an increase of the aspect ratio of the fiber.

Although $k_\textrm{t}$ decreases, the drop is quite slight and $k_\textrm{t}$  keeps almost unchanged with an increase
of the aspect ratio from 0.2 to 1. Those results are obtained based on the condition that the surface
displacements of the ceramic phase and the polymer phase are the same. This is obviously unreasonable.
When the aspect ratio is from 0.2 to 1, the lateral spatial scale of 1--3 composite is comparatively coarse.
In the actual application, when a pressure is applied on 1--3 composite, the displacements of the ceramic
phase and the polymer phase are totally different as it was revealed by a laser probe measurement of the
displacement of oscillating composite plates \cite{Smith91}. Therefore, it is reasonable to doubt that the
boundary condition claiming that `predefined displacements are applied perpendicularly on $C^+$ plane on all nodes',
which means that the displacements or strains of both ceramic phase and polymer phase are restricted to be equal,
is not suitable for the analysis of 1--3 composite with comparatively coarse lateral spatial scale
(i.e., comparatively large aspect ratio from 0.2 to 1). In order to conform to the actual condition, the
boundary condition should be revised as `a predefined pressure is applied perpendicularly on $C^+$ plane'.
In fact, with an increase of the aspect ratio, $k_\textrm{t}$ will drastically decrease instead of keeping almost unchanged.

\section{Proof in the existing reference}

The theoretical model of 1--3 composite has been developed according to Newnham's parallel
connectivity theory \cite{New78}, in which 1--3 composite is restricted to be an effective homogeneous medium.
In W.A.~Smith's model \cite{Smith91} and Chan's model \cite{Wah89}, it was assumed that the spatial
scale of 1--3 composite is so fine that the composite is validly represented as a homogeneous medium
for the frequencies of interest near the thickness resonance. The corresponding approximation embodied
the picture that the ceramic and polymer moved together in a uniform thickness oscillation.
Thus, the vertical strains were the same in both phases. The homogonous medium assumptions are obviously
not satisfied when the spacing between pillars of 1--3 composite are not so fine. So, the theoretical
model will not fit for the analysis of the 1--3 composite in such a case. Electro-elastic
properties of the 1--3 composite will also be affected by the allocation and arrangement of the pillars.

\begin{figure}[!t]
\centerline{
\includegraphics[width=0.65\textwidth]{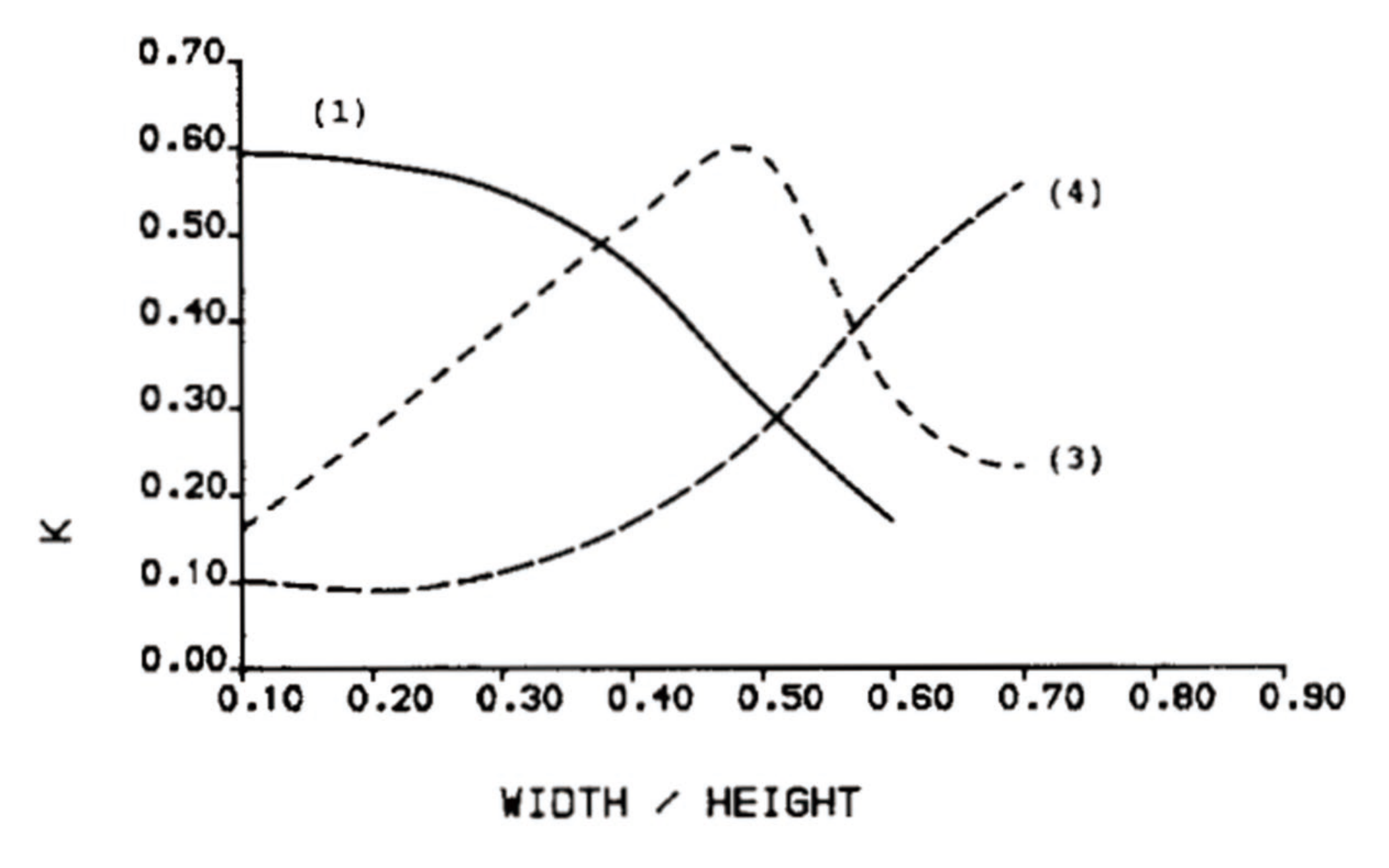}
}
\caption{Dependence of the coupling coefficient in a 20\% circular composite transducer, on the pillar
aspect ratio. Curve (1): thickness mode (concerned). (From reference \cite{Hos91}).} \label{figure-2}
\end{figure}

John A. Hossack and Gordon Hayward \cite{Hos91} thoroughly investigated the effect of ceramic
volume fraction, pillar shape and pillar orientation on the vibrational and electromechanical
characteristics of 1--3 composite transducers using the finite element analysis. It revealed that the
pillar shape and aspect ratio had a great and complex impact on the parameters of the composites.
For the circular section pillars with square arrangement, the pillar aspect ratio is defined as
the ratio of diameter to the height, which is the same as the case in \cite{Mad10}. The two phase
materials are PZT-5A and CY1301-HY1300, whose properties are similar to those of PZT-7A
and Araldite D in reference \cite{Mad10} (from reference \cite{Wah89}), respectively. The dependence
of the coupling coefficient of 20\% circular composite on the pillar aspect ratio is shown in figure~\ref{figure-2}
(figure 11 in \cite{Hos91}). The thickness mode corresponds to the curve (1). It can be seen that
for the 20\%-volume faction composite, it showed a rapid deterioration in the coupling coefficient
for increasing pillar aspect ratios. This is obviously different from the result shown in figure~\ref{figure-1}.

\section{Suggested corrections}

For cylindrical fibers having a square arrangement, the variation of $k_\textrm{t}$ with the aspect ratio
is recalculated under the revised boundary condition claiming that `a predefined pressure is applied
perpendicularly on $C^+$ plane'. The two phase materials are PZT-7A and Araldite D in reference \cite{Mad10}
and the parameters are also taken from reference \cite{Wah89}. (The parameters of reference \cite{Mad10} are
taken from reference \cite{Wah89}). The elastic constants $c_{33}^D$ and $c_{33}^E$ are calculated to obtain the
thickness electromechanical coupling coefficient $k_\textrm{t}=\sqrt{1-c_{33}^E/c_{33}^D}$ \,
($\sqrt{1-c_{33}^E/c_{33}^D}=e_{33}/\sqrt{c_{33}^D\epsilon_{33}^S}=h_{33}/\sqrt{c_{33}^D\beta_{33}^S}$).

To cause the change of the strain in $X_{3-}$ direction, a predefined pressure is loaded
perpendicularly on $C^+$ plane. At the same time, zero displacement is loaded on $C^-$ plane.
Planes $A^+$, $A^-$, $B^+$, $B^-$ are also clamped in their normal directions in order to set
strain tensors in other directions to zero. For $c_{33}^E$, `constant electric
field' means that $C^+$ plane
and $C^-$ plane are short-circuited. Thus, $C^+$ plane and $C^-$ plane are kept at zero potential.
While for $c_{33}^D$, `constant electric displacement field' means that $C^+$ plane and $C^-$ plane are open-circuited.

\begin{figure}[!b]
\centerline{
\includegraphics[width=0.65\textwidth]{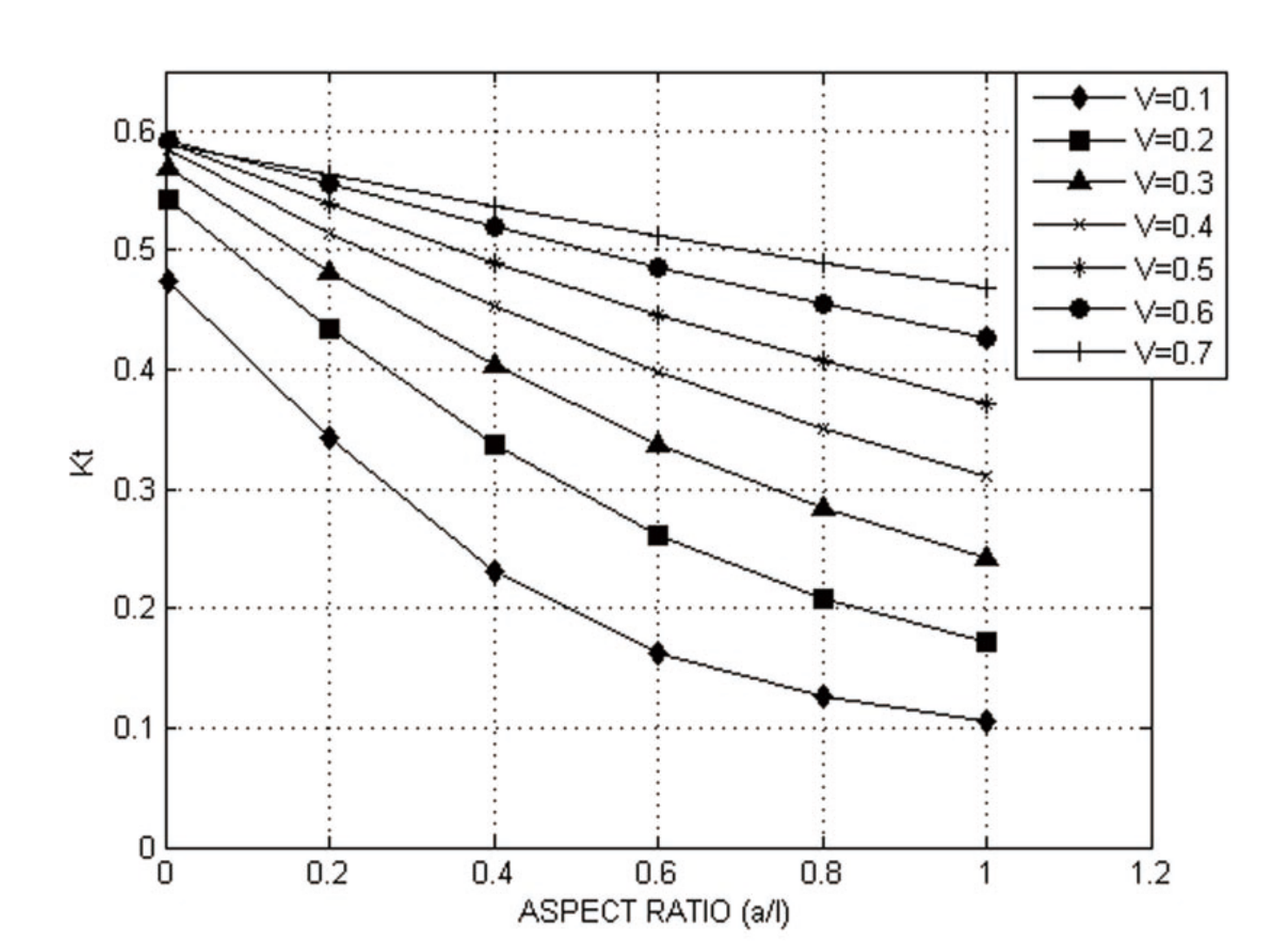}
}
\caption{Suggested correction of variation of $k_\textrm{t}$ with aspect ratio.} \label{figure-3}
\end{figure}

The suggested correction of variation of $k_\textrm{t}$ with the aspect ratio is shown in figure \ref{figure-3}.
Compared with figure \ref{figure-1}, the range of the aspect ratio is extended to $0.005 - 1$.
It can be seen that with an increase of the aspect ratio, $k_\textrm{t}$ drastically decreases, especially
for the low volume fraction.

When the aspect ratio is very small (the lateral spatial scale is very fine),
the values of $k_\textrm{t}$ are comparatively large and are approximately equal to those
in figure \ref{figure-1}. This is because when the lateral spatial scale is fine enough,
the surface displacements of the ceramic phase and the polymer phase are the same under
a pressure load. The effects of boundary conditions claiming that `predefined displacement' and `predefined pressure'
are nearly the same. However, when the aspect ratio is comparatively large (the lateral spatial scale
is comparatively coarse), much of the pressure on the composite will be squandered in compressing
the piezoelectrically inactive phase. The matrix phase presents a displacement with amplitude much
larger than that of the ceramic
phase. Therefore, piezoelectricity of the composite will be greatly
reduced and $k_\textrm{t}$ will drastically decrease.

\ukrainianpart
\title{Коментар до статті С.В. Мадусудгана Рао, Г. Прасад, ``Характеристика  1--3 п'єзоелектричних полімерних
композитів~--- числова і аналітична процедура оцінки вібрацій мод товщин'',
Condensed Matter Physics, 2010, Vol.~13, No.~1, 13703}%

\author{Я. Сан, Ж. Лі}
\address{Вища школа електронної інженерії і комп'ютерних наук, Пекінський університет,
  Пекін, Китай}

\makeukrtitle

\begin{abstract}
\tolerance=3000%
У статті ``Характеристика  1--3 п'єзоелектричних полімерних композитів~--- числова і аналітична процедура
оцінки вібрацій мод товщин'', досліджується залежність коефіцієнта товщини електромеханічного зв'язування
від коефіцієнта волокнистості  п'єзокераміки методом комп'ютерного моделювання скінченного елемента
для різних об'ємних долей п'єзоелектричних волокон в  1--3 композиті. Точність результатів
є під знаком питання, тому що гранична умова, яка зумовлює  зміщення  у напрямку,
перпендикулярному до   $C^+$ площини на всіх нодах, є непридатною для аналізу  1--3 композиту
з порівняльно великим коефіцієнтом від  0.2 до 1.
Тут є представлено обговорення цієї проблеми і запропоновані поправки.

\keywords п'єзоелектричність, п'єзоелектричні матеріали, електромеханічні ефекти, полімерні композити, аналіз скінченного елемента
\end{abstract}

\end{document}